\documentclass[preprint,12pt]{elsarticle}
\usepackage{amsmath}
\usepackage{amsfonts}
\usepackage{amssymb}
\usepackage{graphicx}
\usepackage{hyperref}
\usepackage{clrscode3e}
\usepackage{color}

\hypersetup{hypertex=true,
colorlinks=true,
linkcolor=blue,
anchorcolor=blue,
urlcolor=blue,
citecolor=blue}

\begin{document}

\begin{frontmatter}

\title{Dynamic Moir\'{e}-like pattern in non-Hermitian Wannier-Stark ladder
	system}
\author{H. P. Zhang}
\author{Z. Song}
\ead[Email address: ]{songtc@nankai.edu.cn}
\address{School of Physics, Nankai University, Tianjin 300071, China}

\begin{abstract}
We study the dynamical behavior of the non-Hermitian Wannier-Stark ladder
system, which is a non-Hermitian Su-Schrieffer-Heeger chain with a
position-dependent real potential. In the presence of a linear external
field, we employ the non-Hermitian Floquet method and find that the energy
levels are sensitive to the field. The system exhibits two distinct dynamic
behaviors separated by an exceptional point: one in the $\mathcal{PT}$
symmetrical region associated with two real Wannier-Stark ladders, and
another in the $\mathcal{PT}$\ symmetry-breaking region associated with
complex conjugate ladders. As the boundary between the two regions, two
ladders coalesce into a single ladder. In the case of a non-linear field,
these two distinct regions appear alternately along the chain, exhibiting
dynamic Moir\'{e}-like patterns.
\end{abstract}


\end{frontmatter}


\section{Introduction}

\label{Introduction}

Wannier-Stark (WS) ladders are a theoretical model in solid-state physics
that describe the quantization of electronic states in a crystal under the
influence of a constant electric field \cite
{Bloch1929,Wannier1959,Wannier1960,Glueck2002,Waschke1993}. It is closely
related to Bloch oscillations (BOs) due to the equidistant spectrum. In the
past decade, such a phenomenon has attracted much attention in cold-atoms
physics and photonics due to applications in interferometric measurements
and as a method for manipulating localized wave packets \cite
{Breid2006,Breid2007,Dreisow2009,Kling2010,Ploetz2011}. It can be simulated
using artificial quantum systems, such as superconducting circuits \cite%
{Song2024}. On the other hand, WS ladders are an essential theoretical tool
for understanding and exploring particle transport phenomena in quantum
systems and have potential applications in superconducting quantum computing
and quantum simulation. In the previous study, encompassing both theoretical
and experimental aspects, only a linear external field was considered
because it is central to the formation of an equidistant spectrum.

The main question we address in this work is how a weak non-linear field
influences the dynamics of a WS system. Intuitively, such a field results in
only a slight change in the oscillating frequency along the chain.
Nevertheless, as will be shown, a weak non-linear field leads to a
fascinating phenomenon when a non-Hermitian double-well lattice system is
considered. In this work, we consider a non-Hermitian Su-Schrieffer-Heeger
(SSH) chain with a position-dependent real potential, which is schematically
illustrated in Fig. \ref{fig:1}(a). In the simple case of zero non-Hermitian
hopping ($\beta =0$) and a linear field, the energy levels can be easily
obtained as two sets of WS ladders with $E_{\pm }/\omega =2m\omega $ $\pm 
\sqrt{\left( J/\omega \right) ^{2}+1/4}$, where $m$ is an integer.
Obviously, the energy difference can be zero or $\omega $, corresponding to
two identical or a single WS ladder, respectively, as $\omega $ varies.
Importantly, such a dramatic change can be induced by a shift in the
frequency, $\Delta \omega \approx \omega ^{2}/8$. This indicates that such a
transition is very sensitive to the value of $\omega $, especially in the
limit of small $\omega $. The corresponding distinct dynamical behaviors can
emerge alternatively along the chain when a weak non-linear field is
applied. Furthermore, when $\beta $ is nonzero, the non-Hermitian term can
amplify this phenomenon.

One of the unique features of a non-Hermitian system is the violation of
conservation law of the Dirac probability, based on which, the complex
potential and non-Hermitian hopping strength are employed to describe open
systems phenomenologically \cite{J2004}. Furthermore, unconventional
propagation of light associated with the gain/loss has been demonstrated by
engineering effective non-Hermitian Hamiltonians in optical systems \cite%
{Guo2009,Makris2008,Musslimani2008,Klaiman2008,Longhi2009,ElGanainy2007,Zheng2010}%
.

Exceptional points (EPs), as an exclusive feature of a non-Hermitian system,
are degeneracies of non-Hermitian operators \cite%
{Berry2004,Heiss2012,Miri2019, Zhang2020}. The corresponding eigenstates
coalesce into one state resulting to the incompleteness of Hilbert space.
The peculiar features around EP have sparked tremendous attention to the
classical and quantum photonic systems \cite%
{Doppler2016,Xu2016,Assawaworrarit2017,
Wiersig2014,Wiersig2016,Hodaei2017,Chen2017}. Notably, a coalescing state
has an exclusive feature. Both theoretical and experimental works not only
give an insight into the dynamical property of the non-Hermitian Hamiltonian
but also provide a platform to implement the novel optical phenomenon.

In this work, we study the dynamical behavior of the non-Hermitian WS ladder
system with a weak non-linear field. Analytical analysis and numerical
simulation based on the non-Hermitian Floquet method reveal that a system
with a linear potential exhibits the following features: (i) The spectrum of
the system consists of two sets of complex WS ladders; (ii) they can be two
real WS ladders with identical level spacing, a single coalescing WS ladders
referred to as an EP ladder, and complex conjugate piar of WS ladders; (iii)
As the boundary between the complex and real regions, the EP ladder appears
periodically as the slope of the linear field changes slightly. These allow
for the appearance of dynamic Moir\'{e}-like patterns in the system with a
weak non-linear field.

In general, Moir\'{e} patterns emerge from the superposition of two periodic
structures that have either slightly different periods or different
orientations. These patterns have been realized in materials \cite%
{Yankowitz2012,Ponomarenko2013,Dean2013,Hunt2013,Woods2014,Nakajima2016,Lohse2015}%
. Recently, the influence of Moir\'{e} patterns in physical systems has
attracted considerable interest \cite%
{Ponomarenko2013,Dean2013,Hunt2013,Gorbachev2014,Song2015,Jung2014},
particularly in non-Hermitian systems and the dynamics of such systems \cite%
{Yang2018,Wang2020}. { In
	comparison with existing work on Moir\'{e} patterns, our work has the
	following characteristics: (i) The Moir\'{e}-like patterns presented in this
	work are based on a somewhat different mechanism. They mainly arise from the
	quasi-periodicity resulting from a single periodic structure with a
	non-linear potential, rather than from two periodic structures. (ii) The Moir\'{e}-pattern can be detected
	in the dynamics rather than in static properties. (iii) This is the first
	time we apply the Floquet method to a non-Hermitian system.}

This paper is organized as follows. In Section \ref{Hamiltonian and symmetry}%
, the non-Hermitian SSH model under a position-dependent potential is
introduced. We explore the symmetries of the Hamiltonian with a linear
potential and investigate the possible structures of energy levels based on
these symmetries. In Section \ref{Floquet method} and Section \ref{Phase
diagram}, we employ the non-Hermitian Floquet method to calculate the phase
diagram. In Section \ref{Dynamics Under Linear and Nonlinear Potentials},
the dynamical behavior for the case with a linear potential is studied. We
also propose a model with non-linear potentials and demonstrate the Moir\'{e}%
-like pattern in the dynamics through numerical simulations. A summary is
provided in the Section \ref{Summary}.

\section{Hamiltonian and EP Wannier Stark ladder}

\label{Hamiltonian and symmetry}

\begin{figure*}[tbp]
\centering
\includegraphics[width=\linewidth]{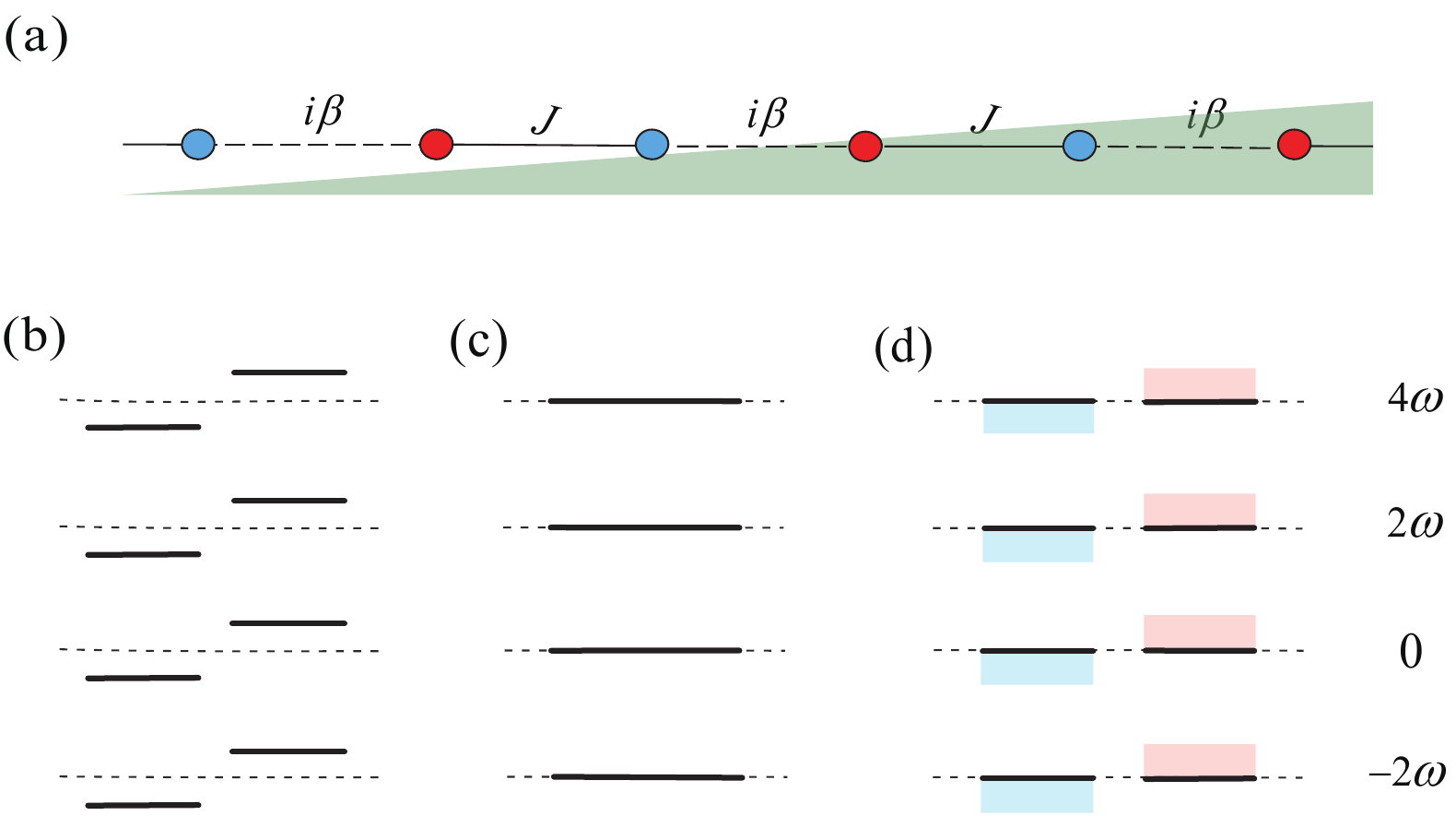}
\caption{Schematic illustrations of (a) the non-Hermitian SSH Hamiltonian
with a tilted field in Eq.\ (\protect\ref{H}), which consists of adjacent
real hopping strength $J$\ and\ imaginary hopping strength $i\protect\beta $
, and (b-d) its consponding energy level structures for three typical cases.
When the on-site tilted potential has a uniform slope $\protect\omega $, the
Hamiltonian obeys the three symmetries defined by Eqs. (\protect\ref{PT}), ( 
\protect\ref{ramped}), and (\protect\ref{modified R}), respectively. These
constrain the energy spectrum partition the energy spectrum into three
distinct categories: (b) two real ladders with an energy shift, forming a
symmetric spectrum with respect to zero energy;\ (c) a single coalescing
ladder with level spacing $2\protect\omega $, (d) two complex conjugate pair
ladders. The black solid lines represent the real part of the energy levels,
while the pink and blue blocks represent the positive imaginary part and
negative imaginary part of the energy levels, respectively.}
\label{fig:1}
\end{figure*}

We consider a non-Hermitian variant of SSH chain \cite%
{Su1980,Takayama1980,Roth1987,Jackiw1976,Heeger1988,Ryu2002,Hasan2010,Qi2011}
imposed by a position dependent field with the Hamiltonian

\begin{equation}
H=\sum\limits_{j=-\infty }^{+\infty }[J(|2j\rangle \langle 2j+1|+\,\mathrm{%
h.c.})+i\beta (|2j\rangle \langle 2j-1|+\,\mathrm{h.c.})]+\sum\limits_{l=-%
\infty }^{+\infty }V_{l}|l\rangle \langle l|]\,,  \label{H}
\end{equation}%
where $|l\rangle $ denotes a site state describing the Wannier state
localized on the $l$th period of the potential. Here, $J$ and $i\beta $ are
the tunneling strength, and $V_{l}$\ a on-site potential. In this paper, we
take $J$ and $\beta >0$\ for the sake of simplicity. The system is
schematically illustrated in Fig.\ \ref{fig:1}(a). In previous work, it has
been shown that the energy levels obeys a simple structures in its real and
imaginary parts when taking the linear potential, $V_{l}=\omega (l+\frac{1}{2%
})$ \cite{Zhang2024}. In this work, we consider the case involving a
non-linear potential $V_{l}$, which will be demonstrated to give rise to a
dynamic Moir\'{e}-like pattern. 

To proceed, we start with the case with\ $V_{l}=\omega (l+\frac{1}{2})$. The
structure of energy levels of $H$ can be obtained from its special
symmetries without the necessity for detailed calculations. We note that the
Hamiltonian has $\mathcal{PT}$ symmetry, satisfying%
\begin{equation}
\mathcal{PT}H\left( \mathcal{PT}\right) ^{-1}=H,  \label{PT}
\end{equation}%
where the linear operator $\mathcal{P}$\ and antilinear operator $\mathcal{T}
$\ are defined as%
\begin{equation}
\left\{ 
\begin{array}{c}
\mathcal{P}|2{l}\mathcal{\rangle }=(-1)^{l}|2l\mathcal{\mathcal{\rangle }}
\\ 
\mathcal{P}|2l+1\mathcal{\mathcal{\rangle }}=(-1)^{l}|2l+1\mathcal{\mathcal{%
\ \rangle }}%
\end{array}
\right. ,\,
\end{equation}%
and $\mathcal{T}i\mathcal{T}^{-1}=-i$. According to non-Hermitian quantum
mechanics \cite{Scholtz1992}, a $\mathcal{PT}$-symmetric Hamiltonian, as a
pseudo-Hermitian system, is characterized by energy levels that appear
either real or in complex conjugate pairs. The eigenstates corresponding to
complex conjugate energy levels are interconnected by the $\mathcal{PT}$
operation. Introducing the translation operator $T_{2}$\ defined as%
\begin{equation}
T_{2}|l\rangle =|l+2\rangle ,
\end{equation}%
the special structure of the linear potential ensures that $H$\ obeys 
\begin{equation}
T_{2}HT_{2}^{-1}=H-2\omega ,  \label{ramped}
\end{equation}%
which is referred to as a ramped translational symmetry. Suppose we have a
solution $|\psi _{0}^{+}\rangle $ of the Schrodinger equation corresponding
to energy $E_{0}^{+}$ 
\begin{equation}
H|\psi _{0}^{+}\rangle =E_{0}^{+}|\psi _{0}^{+}\rangle ,
\end{equation}%
a set of eigenstates can be generated as 
\begin{equation}
|\psi _{n}^{+}\rangle =\left( T_{2}\right) ^{n}|\psi _{0}^{+}\rangle
\end{equation}%
with eigen energy $E_{n}^{+}=E_{0}^{+}+2n\mathcal{\omega }$,\ $(n=0,\pm
1,\pm 2,...)$, which covers half of the complete set eigenstates. In
paralell, one can construct another set of eigenstates 
\begin{equation}
|\psi _{n}^{-}\rangle =\left( T_{2}\right) ^{n}|\psi _{0}^{-}\rangle
\end{equation}%
with eigen energy $E_{n}^{-}=E_{0}^{-}+2n\mathcal{\omega }$,\ $(n=0,\pm
1,\pm 2,...)$, by starting an eigenstate $|\psi _{0}^{-}\rangle $. Without
loss of generality, we can assume that the real part of the energy
eigenvalues, $\left\vert \mathrm{Re}(E_{0}^{\pm })\right\vert $, is less
than or equal to $\mathcal{\omega }$.

We find that the structure of the entire energy spectrum is solely
determined by the base energy level $E_{0}^{\pm }$. As a pseudo-Hermitian
system, it is straightforward to conclude that its eigenvalues $E_{0}^{+}$\
and $E_{0}^{-}$\ are either real or form complex conjugate pair.
Additionally, there exists another symmetry within the system that provides
a constraint for the relationship between $E_{0}^{+}$\ and $E_{0}^{-}$. In
fact, introducing a modified reflection transformation with respect of the
position $m-1/2$,\ 
\begin{equation}
R_{m}|l\rangle =(-1)^{l}|2m-1-l\rangle ,
\end{equation}%
we have 
\begin{equation}
R_{m}HR_{m}^{-1}=-H+2m\omega .  \label{modified R}
\end{equation}%
Applying the above relation on the state $|\psi _{n}^{\pm }\rangle $, we have%
\begin{equation}
HR_{m}|\psi _{n}^{\pm }\rangle =\left( 2m\omega -E_{n}^{\pm }\right)
R_{m}|\psi _{n}^{\pm }\rangle .
\end{equation}%
Specifically, taking $m=0$, we have%
\begin{equation}
H\left( g_{0}|\psi _{0}^{\pm }\rangle \right) =-E_{0}^{\pm }\left(
g_{0}|\psi _{0}^{\pm }\rangle \right) ,
\end{equation}%
\ which results in nothing but 
\begin{equation}
E_{0}^{+}=-E_{0}^{-}.
\end{equation}%
Based on the above analysis, we conclude that: (i) When $E_{0}^{+}$ is
complex, $E_{0}^{\pm }$\ must be imaginary, and the two sets of eigenstates
can be generated as 
\begin{equation}
|\psi _{n}^{+}\rangle =\left( T_{2}\right) ^{n}|\psi _{0}^{+}\rangle = 
\mathcal{PT}|\psi _{n}^{-}\rangle .
\end{equation}%
(ii) When $E_{0}^{+}=E_{0}^{-}=0$, the two sets of WS ladder states coalesce
into a single set, that is, $|\psi _{n}^{+}\rangle =|\psi _{n}^{-}\rangle $.
(iii) When $E_{0}^{+}$ is real, the two sets of energy levels are real, with
an energy shift of $2\left\vert E_{0}^{+}\right\vert $, forming a symmetric
spectrum with respect to zero energy.

The structure of energy levels for each case is schematically illustrated in
Fig.\ \ref{fig:1}(b-d). In the following sections, we will confirm the
aforementioned predictions and provide a detailed solution by introducing
the Floquet method. We will demonstrate that the EP ladder, which serves as
the boundary between complex and real regions, emerges periodically with
minor variations in the slope of the linear field. These allow for the
appearance of dynamic Moir\'{e}-like patterns in the system with a weak
non-linear field.

\section{Non-Hermitian Floquet method}

\label{Floquet method}

In this section, we investigate the energy level structure within the
framework of the non-Hermitian Floquet method, which represents an extension
of the conventional Floquet method \cite{Maksimov2015} to the presnt
non-Hermitian Hamiltonian.

Setting 
\begin{equation}
|\psi \rangle =\sum\limits_{l=-\infty }^{+\infty }(\phi _{l}^{A}|2l\rangle
+\phi _{l}^{B}|2l+1\rangle )
\end{equation}%
the Schrodinger equation $H|\psi \rangle =E|\psi \rangle $\ can be written
as the form

\begin{eqnarray}
2\omega (l+\frac{1}{4})\phi _{l}^{A}+J\phi _{l}^{B}+i\beta \phi _{l-1}^{B}
&=&E\phi _{l}^{A},  \label{S eq} \\
2\omega (l+\frac{3}{4})\phi _{l}^{B}+J\phi _{l}^{A}+i\beta \phi _{l+1}^{A}
&=&E\phi _{l}^{B}.  \notag
\end{eqnarray}%
To solve the coupled equations, we introduce the Bloch wave representation
by the Fourier transformation

\begin{equation}
\phi _{k}^{A,B}=\frac{1}{\sqrt{2\pi }}\sum_{l=-\infty }^{+\infty }\phi
_{l}^{A,B}e^{-ikl},
\end{equation}%
with the wave vector $k\in \lbrack -\pi ,\pi )$, which represents the Eqs. (\ref{S eq}) as two ordinary differential equations%
\begin{equation}
i2\omega \frac{\mathrm{d}\mathbf{Y}_{k}}{\mathrm{d}k}=G_{k}\mathbf{Y}_{k}.
\label{k-S eq}
\end{equation}%
Here the vector $\mathbf{Y}_{k}$ is defined as $\mathbf{Y}_{k}=\left( \phi
_{k}^{A},\phi _{k}^{B}\right) ^{\intercal }$ and $G_{k}$ is a $2\times 2$
non-Hermitian matrix 
\begin{equation}
G_{k}=\left( 
\begin{array}{cc}
E-\frac{\omega }{2} & -J-i\beta e^{-ik} \\ 
-J-i\beta e^{ik} & E-\frac{3}{2}\omega%
\end{array}
\right) .  \label{G}
\end{equation}%
Owing to the requirement for the wave function to be single-valued, we focus
solely on periodic solutions that satisfy $\mathbf{Y}_{2\pi +k}=\mathbf{Y}%
_{k}$. This leads to the quantization of the energy $E$ in the Eq. (\ref{G}).

In fact, Eq. (\ref{k-S eq}) can be regarded as a Schroinger equation for a
non-Hermitian time-dependent Hamiltonian $G_{k}$, assuming $k$ as time $t$.
The periodicity of solutions necessitates that the Floquet operator

\begin{equation}
U=\mathcal{T}\exp \left( \frac{1}{2i\omega }\int_{0}^{2\pi }G_{k}\mathrm{d}
k\right) ,
\end{equation}%
has an eigenvalue equal to $1$, where $\mathcal{T}$ is the time-order
operator. To proceed, we focus on the $2\times 2$ matrix\ 

\begin{equation}
U_{0}=\exp (\frac{iE\pi }{\omega })U=\mathcal{T}\exp \left( \frac{1}{
2i\omega }\int_{0}^{2\pi }G_{k}^{0}\mathrm{d}k\right) ,
\end{equation}%
where the matrix $G_{k}^{0}$\ denotes the matrix $G_{k}$\ at $E=0$. Then the
eigen energy $E_{1,2}$ can be obtained from the two eigen values $\lambda
_{1}$ and $\lambda _{2}$ of the matrix $U_{0}$, by the relation%
\begin{equation}
\lambda _{1,2}=\exp (i\pi E_{1,2}/\omega ).
\end{equation}%
For a given pair of ($\lambda _{1}$, $\lambda _{2}$), we always have 
\begin{equation}
\left\{ 
\begin{array}{c}
E_{1}=\varepsilon _{1}+2m\mathcal{\omega } \\ 
E_{2}=\varepsilon _{2}+2n\mathcal{\omega }%
\end{array}
\right. ,
\end{equation}%
$(m,n=0,\pm 1,\pm 2,...)$ with 
\begin{equation}
\lambda _{1,2}=\exp (i\pi \varepsilon _{1,2}/\omega ),
\end{equation}%
where the complex number $\varepsilon _{1,2}$ is a base point by taking $%
\left\vert \func{Re}\left( \varepsilon _{1,2}\right) \right\vert \leq \omega 
$. This result is consistent with the analysis in the previous section.

In general, analytical expressions for $\lambda _{1,2}$ and $\varepsilon
_{1,2}$ cannot be obtained because they involve the time evolution of a
time-dependent system. However, it provides a method for solving the coupled
differential equations, which serves as the foundation for numerical
simulations. Notably, the Floquet method can be employed to investigate the
structure of the energy levels in a non-Hermitian Stark ladder system, which
is more complicated than that in a Hermitian system. In the subsequent
sections, we will delve into the application of this method, exploring its
practical implications and revealing the spectacular features of a
non-Hermitian SW ladder system. 
\begin{figure*}[t]
\centering
\includegraphics[width=\linewidth]{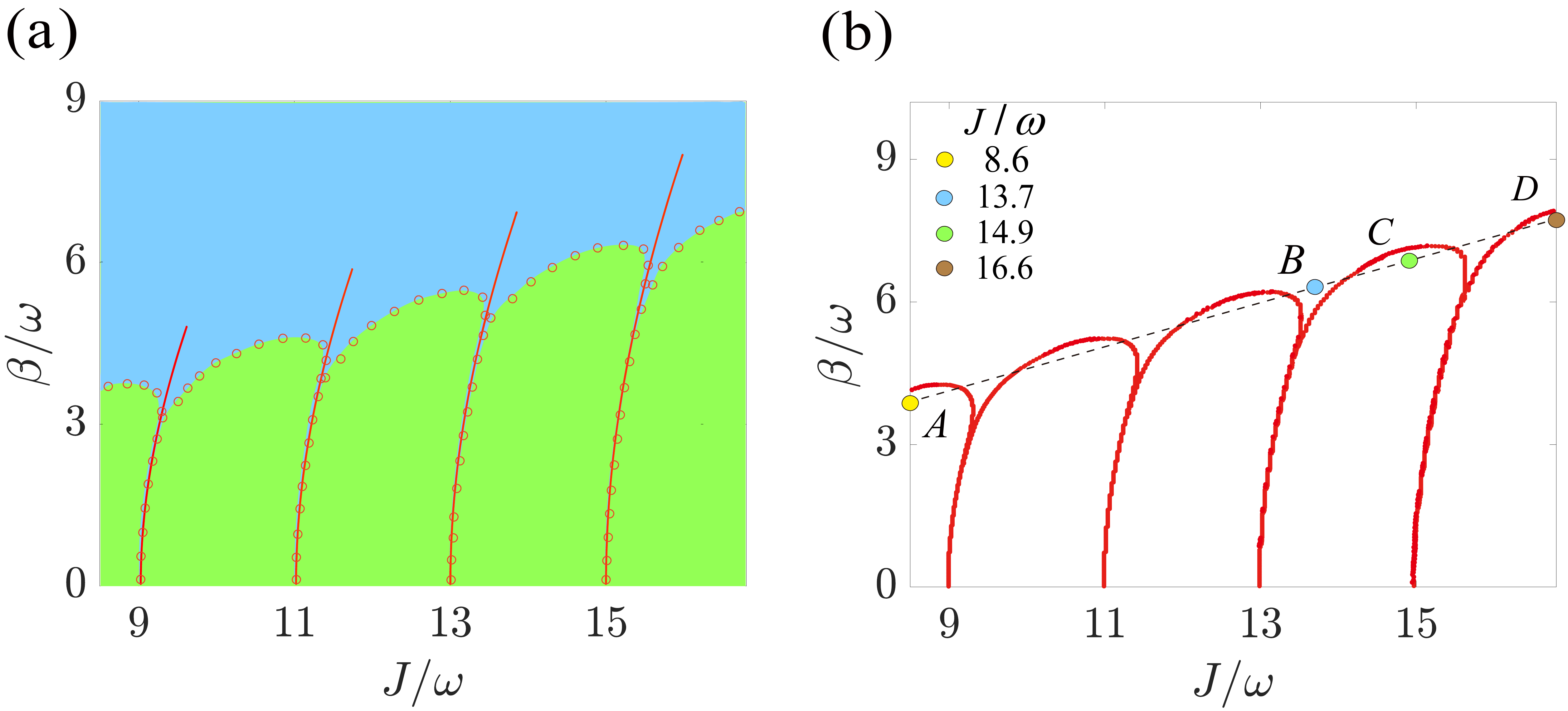}
\caption{(a) The phase diagram of the Hamiltonian, as depicted in Eq.\ ( 
\protect\ref{H}), illustrating the different phases of the system. The green
and blue areas correspond to the parameter ranges where the real
non-degenerate energy levels in Fig.\ \protect\ref{fig:1}(b) and the complex
conjugate pairs of energy levels in Fig.\ \protect\ref{fig:1}(d) occur,
respectively. The red circles represent the distribution of the EP boundary
between phases, while the red solid line is the result predicted by the Eq.\
(\protect\ref{omega}). (b) The same phase diagram as (a) is presented with
an additional specific line, and four typical points are labeled. The dotted
line represents the parameters with $\protect\beta =0.46J$, varying $J/ 
\protect\omega $ from $8.6$ (point A) to $16.6$ (point D). This dotted line
goes through two distinct phases, indicating the transitions from one state
to another periodically. The dynamic behaviors for the systems with
parameters corresponding to points C and B are presented in Figs.\ \protect
\ref{fig:3}(a) and \protect\ref{fig:3}(b), respectively. { This line corresponds to the parameter trajectory in Fig.\ \ref{fig:4}(a), and provides
a way to demonstrate different dynamic behaviors along a chain with the
non-linear potential given by Eq. (\protect\ref{non-linear V}),\ exhibiting
Moir\'{e}-like patterns as shown in Fig.\ \ref{fig:4}(b).} }
\label{fig:2}
\end{figure*}

\section{Phase diagram}

\label{Phase diagram}

\subsection{Phase diagram}

\label{Numerical phase diagram}

Now, we turn to investigate the further relationship between $\varepsilon
_{1,2}$ and $\lambda _{1,2}$.\ The determinant of matrix $U_{0}$\ can be
obtained as

\begin{equation}
\det |U_{0}|=\mathcal{T}\exp [\frac{1}{ 2i\omega }\int_{0}^{2\pi }\text{tr}%
(G_{k}^{0})\mathrm{d}k]=1,
\end{equation}%
which leads to $\lambda _{1}\lambda _{2}=\exp [i\pi (\varepsilon
_{1}+\varepsilon _{2})/\omega ]=1$, or 
\begin{equation}
\varepsilon _{1}+\varepsilon _{2}=0.
\end{equation}%
There exist three typical cases: (i) In the case with $\lambda _{1}=\lambda
_{2}^{\ast }=e^{iv\pi }$ with real $v$, we have $\varepsilon
_{1}=-\varepsilon _{2}=v\omega $, and then two sets of real WS ladders. (ii)
In the case with $\lambda _{1}=\lambda _{2}^{-1}=e^{v\pi }$, we have $%
\varepsilon _{1}=-\varepsilon _{2}=-iv\omega $, and then two sets of WS
ladders, which are all mutually conjugate. (iii) In the case with $\lambda
_{1}=\lambda _{2}=1$, we have $\varepsilon _{1}=\varepsilon _{2}=0$, and
then single set of coalescing ladder. We demonstrate this by the following
simple examples. Considering $J=0$, $\beta \neq 0$, and $\omega \neq 0$, the
system simplifies to a set of independent non-Hermitian dimers. It is
evident that the energy levels are composed of two real or complex Stark
ladders, or they may coalesce into a single ladder. Now we investigate it
within the context of the Floquet method. The corresponding matrix $G^0_{k}$%
\ is given by

\begin{equation}
G^{0}_{k}=-\left( 
\begin{array}{cc}
\frac{\omega }{2} & i\beta e^{-ik} \\ 
i\beta e^{ik} & \frac{3}{2}\omega%
\end{array}
\right) ,
\end{equation}%
where the phases $e^{\pm ik}$ can be eliminated without changing the
physics.\ Then the matrix $U_{0}$\ is obtained as%
\begin{equation}
U_{0}=\cos \Lambda +\frac{i\pi \sin \Lambda }{\Lambda }\left( 
\begin{array}{cc}
\frac{\omega }{2} & \frac{i\beta }{\omega } \\ 
\frac{i\beta }{\omega } & -\frac{\omega }{2}%
\end{array}
\right) ,
\end{equation}%
with $\Lambda =\pi \sqrt{\frac{1}{4}-\left( \frac{\beta }{\omega }\right)
^{2}}$. The eigenvalues are given by 
\begin{equation}
\lambda _{1}=\lambda _{2}^{-1}=\exp (i\Lambda ),
\end{equation}%
which indicates the following scenarios: (i) there are two sets of real
ladders when $\beta <\omega /2$; (ii) there are two sets of complex ladders
when $\beta >\omega /2$; and (iii) there is a single set of coalescing
ladders when $\beta =\omega /2$. Furthermore, these examples demonstrate
that the distinction between a degenerate ladder and a coalescing ladder
stems from whether their corresponding matrix $U_{0}$ is a scalar matrix or
a Jordan block.

For a system with a broad range of parameters $(J,\beta ,\omega )$, the
types of Stark ladders in an infinite chain can be ascertained through
numerical simulations employing the Floquet method. In Fig.\ \ref{fig:2}(a)
we plot the phase diagram in the $J/\omega -\beta /\omega $\ plane. { For a clearer demonstration, we have illustrated in Fig.\ \ref{fig:4}(a) how the real and imaginary parts of the base energy level \( E_{0}^{+} \) change along the trajectories of the two phases in the parameter space.
The parameter trajectory is chosen as the line segment from point A to point D in Fig. \ref{fig:2}(b), and it can be observed that \( E_{0}^{+} \) oscillates between the two phases as the parameter varies.}

\subsection{The distribution of EPs under the weak-field approximation}

\label{The Distribution of EPs}

The numerical results show that for fixed $(J,\beta )$\ the position of EP
strongly depends on the value of $\omega $. This is crucial for the present
work when we consider the case where the external potential $V_{l}$\ is not
linear. In the following we will establish the relation among the parameter $%
(J,\beta ,\omega )$ at EP based on the weak-field approximation.

We rewrite the matrix $G_{k}$ in Eq.\ (\ref{G}) in the form $%
G_{k}=h_{0}+h^{\prime }$, where\ 
\begin{equation}
h_{0}=\left( 
\begin{array}{cc}
E-\frac{\omega }{2} & 0 \\ 
0 & E-\frac{3}{2}\omega%
\end{array}
\right)
\end{equation}%
is in the diagonal form, while $h^{\prime }$\ is given by%
\begin{equation}
h^{\prime }=\left( 
\begin{array}{cc}
0 & -J-i\beta e^{-ik} \\ 
-J-i\beta e^{ik} & 0%
\end{array}
\right) .
\end{equation}%
In the spirit of the interaction picture in quantum mechanics, the Floquet
operator can be expressed as

\begin{eqnarray}
U &=&\mathcal{T}\exp \left( \frac{1}{2i\omega }\int_{0}^{2\pi }(h_{0}+h_{%
\mathrm{I}})\mathrm{d}k\right)  \notag \\
&=&i\exp (-i\frac{\pi E}{\omega })\sigma _{z}\mathcal{T}\exp \left( \frac{1}{%
2i\omega }\int_{0}^{2\pi }h_{\mathrm{I}}\mathrm{d}k\right) ,
\end{eqnarray}%
where%
\begin{equation}
h_{\mathrm{I}}=e^{-ih_{0}k}h^{\prime }e^{ih_{0}k}=-\left( 
\begin{array}{cc}
0 & Je^{ik/2}+i\beta e^{-ik/2} \\ 
Je^{-ik/2}+i\beta e^{ik/2} & 0%
\end{array}%
\right) .
\end{equation}%
Then the calculation of $\mathcal{T}\exp \left( \frac{1}{2i\omega }%
\int_{0}^{2\pi }h_{\mathrm{I}}\mathrm{d}k\right) $\ can be implemented by
the time evolution of the equivalent Schrodinger equation%
\begin{equation}
i\frac{\mathrm{d}\mathbf{\Phi }(t)}{\mathrm{d}t}=-\left( 
\begin{array}{cc}
0 & Je^{i\omega t}+i\beta e^{-i\omega t} \\ 
Je^{-i\omega t}+i\beta e^{i\omega t} & 0%
\end{array}%
\right) \mathbf{\Phi }(t),
\end{equation}%
where $t=k/\left( 2\omega \right) $ can be regarded as time, while $\omega $%
\ is the frequency of the varying Hamiltonian. It is still not exactly
solvable for a given $\omega $. However, an adiabatic solution is available
when the system varies in the limit $\omega \rightarrow 0$. For small $%
\omega $, the adiabatic approximation corresponds to the weak-field
approximation.

For an adiabatic\ process, the evolved state of an initial eigenstate is the
simultaneous eigenstate of the time-dependent Hamiltonian. Specifically, we
have%
\begin{equation}
\mathbf{\Phi }(\pi /\omega )=-\exp i(\varphi _{\mathrm{D}}+\varphi _{\mathrm{B}})\sigma _{z}\mathbf{\Phi }(0),
\end{equation}%
where $\mathbf{\Phi }(0)$ can be one of the eigenstates of the initial
Hamiltonian. The constraint condition given by%
\begin{equation}
U=-i\exp (-i\frac{\pi E}{\omega })\exp i(\varphi _{\mathrm{D}}+\varphi _{ 
\mathrm{B}})=1,
\end{equation}%
requires that

\begin{equation}
\pi E/\omega +\frac{\pi }{2}=\varphi _{\mathrm{D}}+\varphi _{\mathrm{B}}.
\label{E-w}
\end{equation}

A straightforward derivation shows that%
\begin{equation}
\varphi _{\mathrm{D}}=\frac{d}{\omega },\,\varphi _{\mathrm{B}}=-\frac{\pi }{
2},  \notag
\end{equation}%
where $d=\int_{0}^{\pi }\sqrt{J^{2}-\beta ^{2}+i2\beta J\cos \theta }\mathrm{%
\ d}\theta $, which\ can be evaluated by the elliptic integral of the second
kind. Together with the relation $E=2n\omega $\ at EP, we get the EP curve
given by 
\begin{equation}
\omega =\frac{d}{4n\pi +2\pi }.  \label{omega}
\end{equation}%
In Fig.\ \ref{fig:2}(a), we plot the phase diagram in comparison with the
numerical results. It shows that they are consistent with each other for
large values of $\left( J/\omega ,\beta /\omega \right) $. 
\begin{figure*}[t]
\centering
\includegraphics[width=1.07\linewidth]{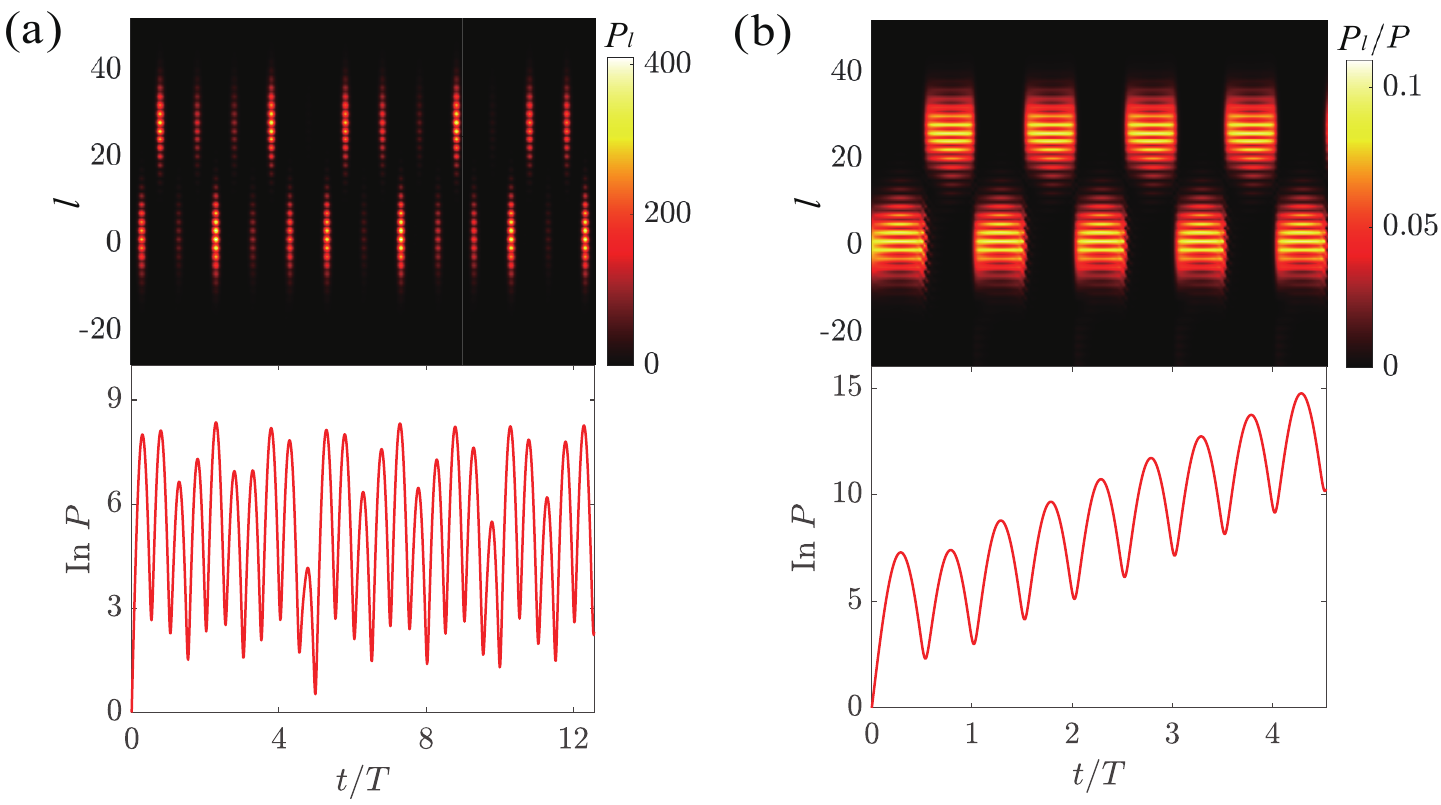}
\caption{The profiles of the evolved states for the initial state given by
Eq.\ (\protect\ref{Gaussian})\ under the systems with parameters
corresponding to points C and B in Fig. \protect\ref{fig:2}(b). 3D plots of
(a) $P_{l}(t)$ and (b) $P_{l}/P$, as defined in Eqs.\ (\protect\ref{P}) and (%
\protect\ref{Pn}), are obtained by exact diagonalization for finite system
with $N=100$.\ We observe that the evolved state (a) is a superposition of
two Bloch oscillations with a period $T=\protect\pi / \protect\omega $, and
(b) represents the amplified Bloch oscillations as expected. The logarithm
of the total Dirac probability, $\text{In}P(t)$, indicates that difference
between two kinds of dynamic behaviors is evident.}
\label{fig:3}
\end{figure*}

\section{Dynamics Under Linear and Nonlinear Potentials}

\label{Dynamics Under Linear and Nonlinear Potentials}

Now we turn to the dynamic characteristics of the phase diagram, starting
with the equations%
\begin{equation}
\left\{ 
\begin{array}{c}
H|\psi _{n}^{+}\rangle =\left( E_{0}^{+}+2n\mathcal{\omega }\right) |\psi
_{n}^{+}\rangle \\ 
H|\psi _{m}^{-}\rangle =\left( -E_{0}^{+}+2m\mathcal{\omega }\right) |\psi
_{m}^{-}\rangle%
\end{array}
\right. ,
\end{equation}%
which govern the dynamics of a given initial state for the case with linear
potential. The dynamics under linear potentials with different slopes
correspond to different points in the phase diagram. Here $E_{0}^{+}$\ is
either real or imaginary. We set $\func{Re}E_{0}^{+}>0$\ or $\func{Im}%
E_{0}^{+}>0$\ for the sake of simplicity.\ Considering an arbitrary initial
state in the form%
\begin{equation}
|\Phi (0)\rangle =\sum_{n,\sigma =\pm }c_{n}^{\sigma }|\psi _{n}^{\sigma
}\rangle ,
\end{equation}%
the time evolution state is%
\begin{equation}
|\Phi (t)\rangle =\sum_{n,\sigma =\pm }e^{-i\sigma \func{Re}E_{0}t}e^{\sigma 
\func{Im}E_{0}t}c_{n}^{\sigma }e^{-i2n\mathcal{\omega }t}|\psi _{n}^{\sigma
}\rangle ,
\end{equation}%
where $\left\{ |\psi _{n}^{\pm }\rangle \right\} $\ is normalized with Dirac
inner product by taking $\left\vert \langle \psi _{n}^{\pm }|\psi _{n}^{\pm
}\rangle \right\vert ^{2}=1$. (i) In the case with $E_{0}^{+}=\left\vert
E_{0}^{+}\right\vert $, it is clear that the evolved state is the
superposition of two Bloch oscillations with a period of $\pi /\omega $. We
are interested in the Dirac probability of the evolved state, which is
defined by

\begin{equation}
P(t)=\left\vert \langle \Phi (t)|\Phi (t)\rangle \right\vert ^{2}.  \label{P}
\end{equation}%
In this case, $P(t)$\ is bounded function with a constant average. (ii) In
the case with $E_{0}^{+}=i\left\vert E_{0}^{+}\right\vert $, we have%
\begin{equation}
|\Phi (t)\rangle \approx e^{E_{0}^{+}t}\sum_{n}c_{n}^{+}e^{-i2n\mathcal{\omega}t}|\psi _{n}^{+}\rangle ,
\end{equation}%
for large $t$. It exhibits amplified Bloch oscillations, characterized by a
probability that increases exponentially. To demonstrate this point, we
perform numerical simulations for a system with a linear potential\textbf{.}
We take a Gaussian wavepacket as the initial state, which is in the form 
\begin{equation}
\left\vert \Phi (0)\right\rangle =\frac{1}{\sqrt{\Omega}}\sum
\limits_{l=-N/2}^{N/2}\exp (-0.01l^{2})\left\vert l\right\rangle .
\label{Gaussian}
\end{equation}%
where $\Omega$ is the normalization coefficient. The evolved state is $%
\left\vert \Phi (t)\right\rangle =e^{-iHt}\left\vert \Phi (0)\right\rangle $%
, which can be obtained by numerical simulation for a finite size system.
The time-dependent Dirac probability distribution is 
\begin{equation}
P_{l}(t)=\left\vert \left\langle l\right\vert \Phi (t)\rangle \right\vert
^{2}.  \label{Pn}
\end{equation}%
The numerical results with two typical sets of parameters are plotted in
Fig.\ \ref{fig:3}. As predicted, the evolved state exhibits the
superposition of two Bloch oscillations with a period of $\pi /\omega $ when
the system has a pure real spectrum. In contrast, the evolved state exhibits
amplified Bloch oscillations when the system has a full complex spectrum.
The evident difference between two kinds of dynamic behaviors lies in the
total Dirac probability: the former has a constant average, while the latter
grows exponentially as time increases. 
\begin{figure}[t]
\centering
\includegraphics[width=0.95\linewidth]{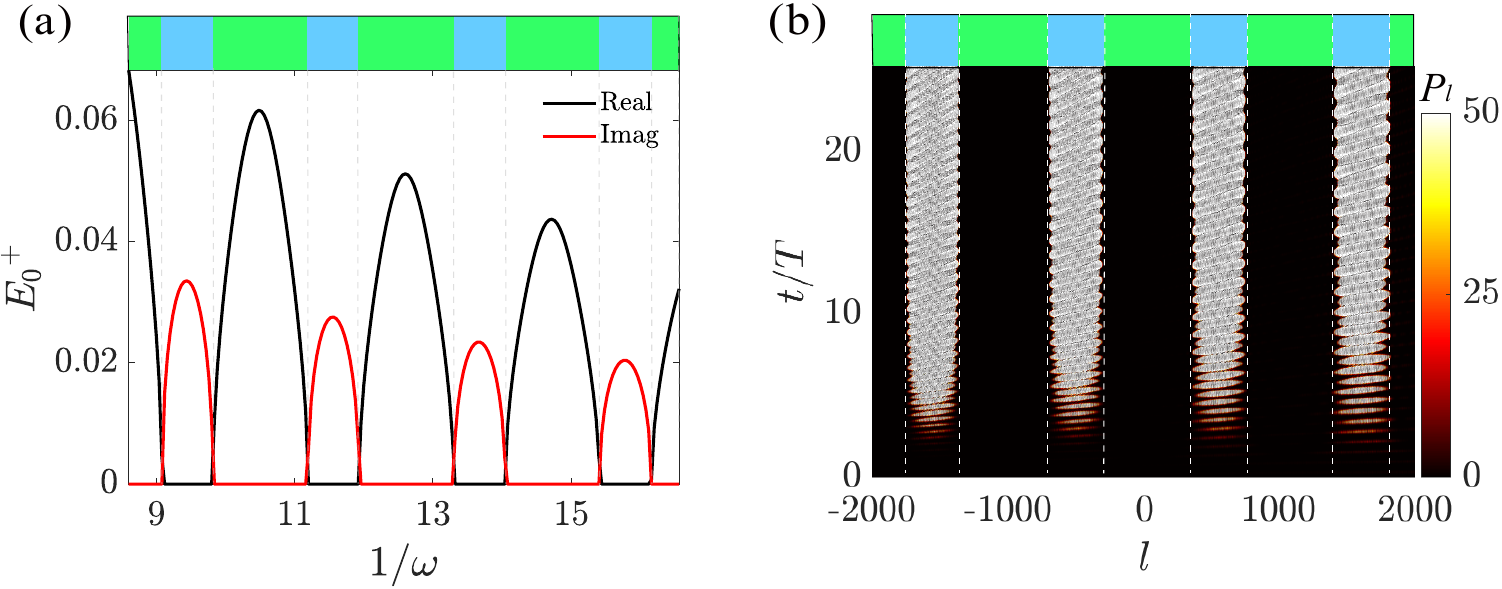}
\caption{(a) The numerical results of the base energy level $E_{0}^+$ along the parameter trajectory of the line segment from point A to point D in Fig.\ \ref{fig:2}(b). The green and blue blocks above represent the $\mathcal{PT}$-symmetric phase and the $\mathcal{PT}$-broken phase, respectively.
	 it can be observed that \( E_{0}^{+} \) oscillates between the two phases as the parameter varies. (b) The profile of the evolved state, given the initial state in Eq.\ (\protect\ref{flat initial state}),\ is computed for the system with $\protect%
\beta =0.46J$\ ($J=1$) and a non-linear potential as described by Eq. (%
\protect\ref{non-linear V}). The 3D plot of $P_{l}(t)$, as defined in Eq. (%
\protect\ref{Pn}), shows that the evolving state exhibits evident Moir\'{e}
-like patterns. Obviously, the dark regions and bright strips correspond exactly to the $\mathcal{PT}$-symmetric phase and the $\mathcal{PT}$-broken phase areas in Figure (a), respectively. The paremeters are $T=\protect\pi \protect\tau =12.6\protect%
\pi $,$\ \protect\gamma =2\times 10^{-3}$, and $N=4000$. }
\label{fig:4}
\end{figure}

We now know that in a system with fixed $J$ and $\beta $, the EP appears as $%
\omega $ varies, which is associated with distinct dynamic behaviors for a
given initial state. It indicates that the EPs occur at different locations
when $\omega $ varies slowly along the chain. Accordingly, the local
dynamics should also be position-dependent. Our primary interest here is in
the effect of slowly varying position-dependent nonlinear potentials on
system dynamics. The potential can map the periodic structure of EPs to the
local dynamics behavior, and induce the formation of a Moir\'{e}-like
pattern. { To demonstrate this point, we perform numerical simulations for a
system with a non-linear potential of the form 
\begin{equation}
V_{l}=\frac{1}{\gamma} \func{In}(\gamma l+\tau),  \label{non-linear V}
\end{equation}%
where the constants $\gamma $ and $\tau $ satisfy $\gamma \ll 1/J\ll \tau $ and $\omega=\text{\textrm{d}}V_l/\text{\textrm{d}}l={1}/(\gamma l+\tau)$ is position-dependent.} It allows mapping the parameter line in Fig. \ref{fig:2}(b)\ onto
an SSH chain, varying the local $J/\omega $ from $8.6$ to $16.6$ along the
chain.\ We take the initial state being distributed on each site with the
equal probability 
\begin{equation}
\left\vert \Phi (0)\right\rangle =\frac{1}{\sqrt{N}}\sum
\limits_{l=-N/2}^{N/2}\left\vert l\right\rangle .  \label{flat initial state}
\end{equation}%
We plot the probability distributions $P_{l}(t)$\ in Fig.\ \ref{fig:4}(b). { The values of 1/$\omega$ in different lattice points of the SSH chain correspond to the horizontal axis of Fig.\ \ref{fig:4}(a). It
can be observed that the evolving state exhibits apparent Moir\'{e}-like
patterns. The dark regions, characterized by a vanishing distribution, are interspersed between the bright strips, which exhibit a large distribution. From Fig.\ \ref{fig:4}(a), it can be seen that these two regions correspond to the non-linear potential falling into two phases with distinct Bloch oscillations: the long-time dynamics will be dominated by the amplified Bloch oscillations, making the distribution in the regions with superimposed Bloch oscillations negligible.}

{ Before ending this paper, we would like to discuss the scheme for
experimentally detecting the {Moir\'{e}-like} pattern in photonic lattices.
We consider a 1D array composed of $2N$ waveguides, which are assumed to be
weakly coupled. According to the coupled-mode theory, light propagation can
be described by a Schroinger-like equation, which is typically used to mimic
the dynamics of a tight-binding system \cite{Xu2016a}. In the tight-binding approximation,
the propagation along the $z$-direction of the optical wave in such a
photonic lattice is described by coupled-mode equations. The propagation of
the optical wave in this photonic lattice is described by the following
coupled-mode equations

\begin{equation}
	\frac{\text{\textrm{d}}A_{l}(z)}{\text{\textrm{d}}z}=i\left( \kappa
	_{l,l-1}A_{l-1}+\kappa _{l,l+1}A_{l+1}\right) ,
\end{equation}%
where $A_{l}(z)$ is the propagating amplitude in the $l$th waveguide, and $%
\kappa _{l,l\pm 1}$ is the coupling coefficient between adjacent waveguides.
To mimic Bloch oscillations, the photonic lattice is bent to modify its
spatial distribution of the optical potential. The curvature is seen as an
inertial force acting on the optical wave. The corresponding equation is%
\begin{equation}
	\frac{\text{\textrm{d}}A_{l}(z)}{\text{\textrm{d}}z}=i\left( \kappa
	_{l,l-1}A_{l-1}+\kappa _{l,l+1}A_{l+1}\right) +iF_{l}lA_{l},
\end{equation}%
where $F_{l}$ is the inertial force acting on the optical wave in the $l$th
waveguide. The key point related to the present topic is as follows. In a
region consisting of a small number of waveguides, $F_{l}$ can be regarded
as approximately $l$-independent. However, the curvature in the radial
direction cannot be regarded as a constant, leading to a nonlinear potential
when a large $N$ is considered. According to our theoretical analysis and
the results of numerical simulations, the shift of $F_{l}$\ across a single period of {Moir\'{e}-like pattern} is estimated as $\{[\mathrm{d}(F_{l}l)/\mathrm{d}%
t]|_{l=l_{0}+\Delta N}\}^{-1}-\{[\mathrm{d}(F_{l}l)/\mathrm{d}%
t]|_{l=l_{0}}\}^{-1}\approx 2$, with $\Delta N=1000$, from the expression of 
$F_{l}$, given by

\begin{equation}
	F_{l}=\frac{1}{0.002l}\ln (0.002l+12.6).
\end{equation}%
Based on these {parameters, }the geometry of the sample, including the range
of the bent radius, and the minimum number of waveguides, can be determined. Additionally, in principle, our findings could be simulated in these systems beyond photonic systems, such as the electric circuit system \cite{Zhang20221111} and the superconducting circuit system \cite{Song2024}.}

\section{Summary}

\label{Summary}

In summary, we have investigated the dynamical behavior of the non-Hermitian
SSH chain, focusing on the effects induced by a non-linear field rather than
a linear one. We have shown that, by analytical analysis and numerical
simulation based on the non-Hermitian Floquet method, such a system with a
linear potential exhibits the following features: (i) The spectrum of the
system consists of two sets of complex WS ladders; (ii) they can be two real
WS ladders with identical level spacing, a single coalescing WS ladders
referred to as an EP ladder, and complex conjugate piar of WS ladders; (iii)
As the boundary between the complex and real regions, the EP ladder appears
periodically as the slope of the linear field changes slightly. These allow
for the appearance of dynamic Moir\'{e}-like patterns in the system with a
weak non-linear field. Our findings not only shed light on the dynamical
properties of non-Hermitian Hamiltonians but also offer a platform for
implementing novel optical phenomena.

\section*{Acknowledgment}

This work was supported by National Natural Science Foundation of China
(under Grant No. 12374461). 

\end{document}